\documentclass[twocolumn,aps,showpacs,prl,tightenlines,amsmath,amssymb]{revtex4}
\usepackage{graphicx}
\usepackage{amssymb}
\usepackage{slashed}
\usepackage{dcolumn}
\usepackage{amsmath}
\usepackage{bm}
\usepackage{colordvi}
\usepackage{mathrsfs}
\makeatletter

\newcommand{\Rmnum}[1]{\expandafter\@slowromancap\romannumeral #1@}
\makeatother

\begin{document}

\title{Theory of Higgs Modes in $d$-Wave Superconductors}

\author{F. Yang}
\email{yfgq@mail.ustc.edu.cn.}

\affiliation{Hefei National Laboratory for Physical Sciences at
Microscale, Department of Physics, and CAS Key Laboratory of Strongly-Coupled
Quantum Matter Physics, University of Science and Technology of China, Hefei,
Anhui, 230026, China}

\author{M. W. Wu}
\email{mwwu@ustc.edu.cn.}

\affiliation{Hefei National Laboratory for Physical Sciences at
Microscale, Department of Physics, and CAS Key Laboratory of Strongly-Coupled
Quantum Matter Physics, University of Science and Technology of China, Hefei,
Anhui, 230026, China}

\date{\today}

\begin{abstract}
  
  By applying a microscopic gauge-invariant kinetic theory in $d$-wave
  superconductors, we analytically derive the energy spectra of the breathing
  Higgs mode and, in particular, rotating Higgs mode that is unique for the
  $d$-wave order parameter. Analytical investigation on their dynamic
  properties is also revealed.
  We show that the breathing Higgs mode is optically
  visible in the second-order regime. Whereas the rotating Higgs
  mode is optically inactive, we show that this mode can be
  detected in the pseudogap phase by magnetic resonance experiment. It
  is interesting to find that with a longitudinal temperature gradient, 
  the charge-neutral rotating Higgs mode generates a
  thermal Hall current by magnetic field in the pseudogap phase, providing a unique scheme for its detection.

\end{abstract}


\maketitle 

{\em Introduction.---}In the past few decades, collective excitation
in the field of superconductivity has attracted much attention.  
Due to the angular and radial excitations in the Mexican-hat
potential of free energy~\cite{Am0}, two types of the collective excitations emerge:
Nambu-Goldstone~\cite{gi0,AK,Gm1,Gm2,Ba0,pi1} and Higgs~\cite{OD1,OD2,OD3,Am12} modes, 
which describe the phase and amplitude fluctuations of the superconducting
order parameter respectively. These modes are decoupled in equilibrium
state as they represent mutually orthogonal excitations~\cite{Am0}.  
The Nambu-Goldstone mode corresponds to
gapless Goldstone boson due to the spontaneous breaking of
continuous $U(1)$ symmetry by order parameter~\cite{Gm1,Gm2}.
For the Higgs mode, theoretical studies in conventional $s$-wave
superconductors reveal an excitation gap at long wavelength,
twice of the superconducting gap~\cite{Am0,OD1}. Whereas being
charge neutral and spinless, this mode has long been experimentally
elusive. Until recently, thanks to advanced ultrafast terahertz pump-probe
technique, the Higgs mode in conventional
superconductors has been observed from the optically excited
oscillation of the superfluid density in second-order regime~\cite{NL1,NL2,NL3,NL4}. The 
most convincing evidence comes from the triggered resonance at
the optical frequency which equals to superconducting gap~\cite{NL2,NL3,NL4}.
The Higgs mode has since stimulated a lot of
experimental interest in a variety of contexts,
extending to unconventional superconductors.
Particularly, recent experiments in 
cuprate superconductor reported similar oscillation of 
superfluid density in the second-order optical
response~\cite{DHM1,DHM2,DHM3}, indicating the observation of the 
Higgs mode in high-$T_c$ superconductors.        

Despite experimental activity, the theoretical description of the 
Higgs mode in high-$T_c$ superconductors remains wide open~\cite{DHMT1,DHMT2}.
In high-$T_c$ superconductors, a finite amplitude of the order
parameter persists up to $T^*$, which is well above
$T_c$~\cite{PG1,PG2,PG3,PG4,PG5}, indicating 
the existence of the Higgs mode not only in superconducting phase
but also in pseudogap one. Moreover, compared with the $s$-wave
case, the high-$T_c$ 
superconductivity with the $d$-wave order parameter and hence the lower
rotating symmetry supports the additional Higgs modes.
Early symmetry analysis revealed the existence of the breathing and
rotating Higgs modes~\cite{DHMT1}, 
which for the case of $d_{x^2-y^2}$-wave order parameter correspond to
$d_{x^2-y^2}$-wave and $d_{xy}$-wave amplitude fluctuations, respectively.
The breathing Higgs mode is expected to be similar to
conventional $s$-wave Higgs mode, whereas the rotating one is unique.  
Recent experiment in cuprate superconductor has
reported that the direction of the order parameter detaches
from that of lattice~\cite{Ro}. A detectable response of 
rotating Higgs mode to the external probe is therefore expected.
Nevertheless, the energy spectra and dynamic
properties for both Higgs modes in $d$-wave superconductors are still
unclear in the literature. Considering the growing experimental
findings, a full theoretical investigation on these modes becomes imperative.    

In this Letter, we use the microscopic gauge-invariant
kinetic equation (GIKE) approach~\cite{GOBE1,GOBE2} to investigate the Higgs modes for $d$-wave order parameter. For the first time, we provide the analytic expressions for 
energy spectra of both breathing and rotating Higgs modes.  
Then, investigation on their dynamic
properties is presented. The breathing Higgs mode is optically
visible in the second-order regime, 
irrelevant of the optical polarization direction. 
Whereas the rotating Higgs mode is optically inactive,
we show that this mode responds to magnetic field in
the linear regime, suggesting a possible detection by magnetic resonance
experiment in the pseudogap phase. It is interesting to find that the
charge-neutral rotating Higgs mode, which does not manifest itself in
the electric measurement, generates a thermal Hall current by magnetic
field in the presence of temperature gradient in pseudogap phase, providing a unique scheme for its detection.   

{\em Hamiltonian.---}We begin our analysis with a free two-dimensional
BCS-like Hamiltonian  
\begin{equation}
\label{BdG}
H=\sum_{{\bf k},s=\uparrow,\downarrow}\xi_{\bf k}c^{\dagger}_{{\bf k}s}c_{{\bf
    k}s}-\sum_{{\bf kk'}}g_{\bf
  kk'}c^{\dagger}_{{\bf k}\uparrow}c^{\dagger}_{-{\bf
    k}\downarrow}c_{-{\bf k'}\downarrow}c_{{\bf k'}\uparrow},
\end{equation}
where  $\xi_{k}=\varepsilon_k-\mu$
and $\varepsilon_k={k^2}/{(2m)}$ with
$m$ and $\mu$ being the effective mass and chemical potential, respectively;
$g_{\bf kk'}$ denotes the
pairing interaction. The order parameter is determined by
{\small $\Delta_{{\bf k}0}=-\sum_{\bf k'}g_{\bf kk'}\langle{c_{-{\bf
        k'}\downarrow}c_{{\bf k'}\uparrow}}\rangle$}. This leads to  
the Bogoliubov quasiparticle energy {\small $E_{\bf k}=\sqrt{\xi_k^2+\Delta^2_{{\bf k}0}}$}. 
In the present analysis,    
we approximately take the pairing interaction $g_{\bf kk'}$ around the Fermi surface (i.e., $k=k'=k_F$) so that the order parameter only has angular dependence of the momentum. 
In the presence of the translational and time-reversal symmetries~\cite{SM1}, the
pairing interaction {\small $g_{\bf kk'}=\sum_{m=0}D_m\cos{m}(\theta_{\bf k}-\theta_{\bf k'})$}
where $\theta_{\bf k}$ denotes the angle of ${\bf k}$ vector.
For $d$-wave superconductivity,
$D_2$ channel dominates, i.e.,
\begin{equation}\label{D2}
g_{\bf kk'}{\approx}D_2\cos{2}(\theta_{\bf k}-\theta_{\bf k'}).  
\end{equation}
However, $g_{\bf kk'}$ here exhibits the rotational symmetry of the
polar axis in momentum space, 
and thus, can not determine the direction of the order
parameter. This direction, as recent experiment
reported~\cite{Ro}, also
detaches from that of lattice. The practical formation of
the $d$-wave order parameter with a certain direction
thus spontaneously breaks the rotational
symmetry. Without losing
generality, we choose $d_{x^2-y^2}$ order parameter for analysis,
i.e., $\Delta_{{\bf k}0}=\Delta_0\cos2\theta_{\bf k}$.
As this finite order parameter in high-$T_c$ superconductors persists
up to $T^*$, which is well above $T_c$~\cite{PG1,PG2,PG3,PG4,PG5}, our
calculation of its amplitude fluctuation in the following 
can be applied not only in superconducting phase but also in
pseudogap one. 

{\em GIKE approach.---}To calculate the Higgs mode, we 
use the GIKE approach~\cite{GOBE1,GOBE2}, which has been successfully
applied to study collective modes and their electromagnetic properties
in conventional superconductors~\cite{GOBE2}. In this microscopic approach,
the response of system is described by 
density matrix $\rho_{\bf k}=\rho^{(0)}_{\bf k}+\delta\rho_{\bf k}(R)$ in
Nambu space, which consists of the equilibrium part {\small
  $\rho^{(0)}_{\bf k}=\frac{1}{2}-\frac{1-2f(E_{\bf
      k})}{2}(\frac{\xi_k}{E_{\bf k}}\tau_3+\frac{\Delta_{{\bf
        k}0}}{E_{\bf k}}\tau_1)$} and nonequilibrium one {\small $\delta\rho_{\bf k}(R)$}.
Here, {\small $f(x)=\frac{1}{e^{x/T}+1}$} denotes the
Fermi-distribution function with $T$ being the temperature;
$\tau_i$ are Pauli matrices in particle-hole space; {\small $R=(t,{\bf R})$} represents the
center-of-mass (CM) coordinate. The off-diagonal
components of $\rho_{\bf k}$ self-consistently determine the order parameter:
\begin{eqnarray}
{\hat \Delta_{\bf k}({R})}&=&[\Delta_{{\bf k}0}+\delta\Delta_{\bf
      k}(R)]e^{i\delta\theta(R)}\tau_++h.c.\nonumber\\&=&-\sum_{\bf k'}g_{\bf kk'}(\rho_{{\bf k'}+}\tau_++h.c.),\label{g1}
\end{eqnarray}  
where $\rho_{{\bf k}i}$ stands for the $\tau_i$ component of $\rho_{\bf k}$;
$\delta\Delta_{\bf k}$ and $\delta\theta$ denote the amplitude and
phase fluctuations of the order parameter, respectively.

To determine the nonequilibrium property/fluctuation,
one needs to solve
$\delta\rho_{\bf k}$ from the GIKE
\begin{equation}\label{KE}
  \partial_t\rho_{\bf
    k}+\partial_t\rho_{\bf
    k}\big|_{\rm coh}+\partial_t\rho_{\bf
    k}\big|_{\rm diff}+\partial_t\rho_{\bf
    k}\big|_{\rm dri}=\partial_t\rho_{\bf k}\big|_{\rm scat}.
\end{equation}
The coherent terms are given by 
\begin{equation}
\resizebox{.85\hsize}{!}{$\partial_t\rho_{\bf k}\big|_{\rm coh}=i\Big[\big(\xi_k\!+\!e\phi\!+\!\mu_H\!+\!\frac{e^2A^2}{2m}\big)\tau_3\!+{\hat
      \Delta_{\bf k}({R})},\rho_{\bf
  k}\Big],$}
\end{equation}
where $\phi$ and ${\bf A}$ stand for the scalar and
vector potentials, respectively; 
{\small $\mu_H(R)=\sum_{\bf R'}V_{{\bf R}-{\bf R'}}{\delta}n({R}')$}
describes the induced scalar potential by the density fluctuation
{\small ${\delta}n({R}')$}, with {\small $V_{\bf R-R'}$} being the
Coulomb potential.

The drive terms by electromagnetic fields read
\begin{eqnarray}
  &&\partial_t\rho_{\bf
    k}\big|_{\rm dri}=\resizebox{.03\hsize}{!}{$\frac{1}{2}$}\{e{\bf E}\tau_3-({\bm \nabla}_{\bf R}\!-\!2ie{\bf A}\tau_3){{\hat
  \Delta}_{\bf k}(R)},\partial_{\bf k}\rho_{\bf
k}\}\nonumber\\
&&\mbox{}\!-\!\resizebox{.033\hsize}{!}{$\frac{i}{8}$}[({\bm \nabla_{\bf R}}\!-\!2ie{\bf A}\tau_3)({\bm
    \nabla_{\bf R}}\!-\!2ie{\bf A}\tau_3){\hat
  \Delta}_{\bf k}(R),\partial_{\bf k}\partial_{\bf k}\rho_{\bf
      k}]\nonumber\\
&&\mbox{}\!+\!\resizebox{.125\hsize}{!}{$\frac{{\bf k}\times{e}{\bf B}}{4m}$}[\tau_3,\tau_3\partial_{\bf k}\rho_{\bf
k}]\!+\!\resizebox{.12\hsize}{!}{$\frac{{\bf k}\times{e}{\bf B}}{2m}$}\{\tau_3,\tau_3\partial_{\bf k}\rho_{\bf
k}\},\label{drive}
\end{eqnarray}
with the electric field {\small $e{\bf E}=-{\bm \nabla}_{\bf R}(e\phi+\mu_H)-\partial_te{\bf A}$} and magnetic field {\small ${\bf B}={\bm \nabla}_{\bf
    R}\times{\bf A}$}. By the Meissner effect, the magnetic field {\small
  ${\bf B}$} is expelled from superconductor in 
superconducting phase. Whereas in pseudogap phase
where the superfluid density vanishes, ${\bf B}$ can be applied.
Here, we consider a perpendicular ${\bf B}$ to the conducting layer.

The diffusion terms due to spatial inhomogeneity read
\begin{equation}
\partial_t\resizebox{.1\hsize}{!}{$\rho_{\bf
    k}\big|_{\rm diff}$}\!=\!\{\resizebox{.06\hsize}{!}{$\frac{\bf k}{2m}$}\tau_3,\resizebox{.045\hsize}{!}{${\bm \nabla}_{\bf R}$}\resizebox{.04\hsize}{!}{$\rho_{\bf
    k}$}\}\!-e[\resizebox{.1\hsize}{!}{$\frac{{\nabla_{\bf R}}\circ{\bf A}}{4m}$},\tau_3\resizebox{.04\hsize}{!}{$\rho_{\bf
    k}$}]\!-\!\resizebox{.028\hsize}{!}{$\frac{i}{4}$}[\varepsilon_{\resizebox{.035\hsize}{!}{${\bm \nabla_{\bf R}}$}
}\tau_3,\resizebox{.04\hsize}{!}{$\rho_{\bf
    k}$}],
\end{equation}
with {\small ${\nabla_{\bf R}}\!\circ\!{\bf A}=(2{\bf A}\!\cdot\!{\bm \nabla}_{\bf R}\!+\!{\bm
    \nabla}_{\bf R}\!\cdot\!{\bf A})\tau_3$}. The lengthy scattering terms
$\partial_t\rho_{\bf k}|_{\rm scat}$ are presented in the Supplemental Material [see Eq.~(S7)~\cite{SM}].  

{\em Calculation of Higgs modes.---}The phase fluctuation in
Eq.~(\ref{g1}) enlarges the difficulty to solve 
$\delta\rho_{\bf k}$ from Eq.~(\ref{KE}), but it can be {\em effectively} removed by unitary
transformation {\small $\rho_{\bf
    k}(R)\rightarrow{e^{i\tau_3\delta\theta(R)/2}}\rho_{\bf
    k}(R)e^{-i\tau_3\delta\theta(R)/2}$} [see Supplemental Material, Eq.~(S3)~\cite{SM}]. Then, for weak probe, by 
expanding {\small $\delta\rho_{\bf k}=\delta\rho^{(1)}_{\bf
    k}+\delta\rho^{(2)}_{\bf k}$} with {\small $\delta\rho^{(1)}_{\bf k}$}
and {\small $\delta\rho^{(2)}_{\bf k}$} being the first and second order responses to
external probe, the GIKE becomes a chain of equations, as its first order
only involves {\small $\delta\rho^{(1)}_{\bf k}$} and equilibrium {\small
  $\rho^{(0)}_{\bf k}$} and its second order involves {\small
  $\delta\rho^{(2)}_{\bf k}$}, {\small $\delta\rho^{(1)}_{\bf k}$} and {\small
  $\rho^{(0)}_{\bf k}$}.
Consequently, starting from the lowest order, we calculate {\small $\delta\rho^{(1)}_{\bf
    k}$} and {\small $\delta\rho^{(2)}_{\bf
    k}$} in sequence, whose lengthy expressions are presented in the Supplemental Material [see Eqs.~(S16) and (S28) \cite{SM}].

Then, we can finally determine the Higgs modes. Particularly, Eq.~(\ref{g1}) after above unitary transformation has variation {\small $\Delta_{{\bf k}0}\!+\!\delta\Delta_{\bf k}\!=\!-\sum_{\bf k'}g_{\bf kk'}\rho_{{\bf k'}1}$},
from which by the solved $\delta\rho_{\bf k}$ we can
self-consistently obtain Higgs mode 
\begin{equation}
\delta\Delta_{\bf k}=-\sum_{\bf k'}g_{\bf
  kk'}\delta\rho_{{\bf k'}1}.
\end{equation}
By further considering $g_{\bf kk'}$ in Eq.~(\ref{D2}), 
the Higgs mode
\begin{equation}
\delta\Delta_{\bf
  k}(R)=\delta\Delta_{B}(R)\cos(2\theta_{\bf k})+\delta\Delta_{R}(R)\sin(2\theta_{\bf k}),
\end{equation}
consists of breathing {\small $\delta\Delta_{B}(R)$} and rotating
{\small $\delta\Delta_{R}(R)$} parts.

{\em Breathing Higgs mode.---}In CM frequency and momentum
space [{\small $R=(t,{\bf R})\rightarrow{q}=(\omega,{\bf q})$}], we find the optical
response of the breathing Higgs mode at weak scattering (for details, see Supplemental Material \cite{SM}) 
\begin{widetext}
\begin{equation}
\left(\omega^2_{B}-\omega^2\right)\delta\Delta_{B}(q)=\Big(\frac{e{\bf
      E}_q\cdot{i}{\bf
      q}}{m}\Big){8u_{\omega}\Delta_0}+v^2_F\left[\frac{2}{i\omega_{\rm
  H}}{\overline{\Big(\frac{e{\bf
            E}_q}{i\omega_{\rm P}}\Big)\cdot{e}{\bf
          E}_q}}+2\overline{\Big(\frac{e{\bf 
            E}_q}{i\omega_{\rm P}}\Big)\cdot{e}{\bf A}_q}+e^2\overline{{\bf
        A}^2_q}\right]\Delta_0d_{\omega},  \label{BH}
\end{equation}
\end{widetext}
where  {\small ${\overline{A_qB_q}}=\int{dq'}A_{q'}B_{q-q'}$};
{\small $i\omega_{\rm P}=i\omega+\Gamma_p$} and {\small
  $i\omega_{\rm H}=i\omega+\Gamma_H$},
with {\small $\Gamma_p$ ($\Gamma_H$)} being the relaxation rate of
nonequilibrium {\small $\delta\rho_{{\bf k}0}$ ($\delta\rho_{{\bf
      k}\pm}$)} from the scattering;
the energy spectrum
\begin{equation}
\resizebox{.58\hsize}{!}{$\omega_B=2\Delta_0\sqrt{\frac{\sum_{\bf k}\cos^4(2\theta_{\bf
        k})a_{\bf k}z_{\omega_B,{\bf k}}}{\sum_{\bf k}\cos^2(2\theta_{\bf
        k})a_{\bf k}z_{\omega_B,{\bf k}}}}$},\label{OB}
\end{equation}
with
{\small $a_{\bf k}=\frac{1-f(E_{\bf k})}{2E_{\bf k}}$} and {\small
  $z_{\omega,{\bf k}}=\frac{1}{4E_{\bf k}^2-\omega^2}$}; 
$u_{\omega}$ and $d_{\omega}$ are the dimensionless coefficients 
for the linear and second-order responses [see Supplemental Material, Eqs.~(S31) and~(S32) \cite{SM}], respectively.

In this response function, it is first noted from the left-hand side
that due to the lower rotational symmetry for the $d$-wave
order parameter [i.e., {\small $\cos^4(2\theta_{\bf
      k})\le\cos^2(2\theta_{\bf k})$} in Eq.~(\ref{OB})], the energy
spectrum of its breathing Higgs mode $\omega_B<2\Delta_0$, in
contrast to the $s$-wave case where the Higgs-mode energy appears at
$2\Delta_0$. Particularly, through a numerical calculation of
    Eq.~(\ref{OB}), we find $\omega_B\sim1.7\Delta_0$ for a wide
    range of parameter choice ($T$ and $\Delta_0$). Furthermore, the first term on
the right-hand side of Eq.~(\ref{BH}) denotes the linear response, which vanishes in the
long-wavelength limit ($q=0$) as it should be for the charge-neutral mode. The second term,
i.e., the second-order response, which originates from the drive terms
[Eq.~(\ref{drive})],
is finite at $q=0$, similar
to the $s$-wave case~\cite{GOBE2}.
Particularly, this response is totally irrelevant of
optical polarization direction, consistent with the experimental
finding~\cite{DHM3}. This actually is very natural considering the fact that 
the optical field in the long-wavelength limit is unaware of the relative
momentum of the pairing electrons (i.e., pairing symmetry).

In the optical detection, for the case with the
multi-cycle terahertz pulse which possesses a stable phase as well
as a narrow frequency bandwidth, one approximately has $E_q, A_q\sim\delta(\omega-\Omega)\delta(q)$, and then, from
Eq.~(\ref{BH}), $\delta\Delta_{B}$ in the time domain becomes
\begin{equation}
\resizebox{.87\hsize}{!}{$\delta\Delta_B(t)=\frac{e^2v^2_F\left[{A}^2_{\Omega}+\frac{2{E}_{\Omega}A_{\Omega}}{i\Omega_{\rm
      P}}+\frac{2E^2_{\Omega}}{i\Omega_{\rm
        P}(i2\Omega+\Gamma_{\rm
      H})}\right]d_{2\Omega}\Delta_0e^{2i\Omega{t}}}{\omega^2_{B}-(2\Omega)^2}$},
\end{equation}
which oscillates at twice optical frequency and exhibits a
resonance at $2\Omega=\omega_B$, similar to the $s$-wave case but with $\omega_B<2\Delta_0$.
As for the case with a short terahertz pulse which possesses the broad
frequency bandwidth, no pole emerges from the electromagnetic fields
in Eq.~(\ref{BH}).
In this circumstance, by neglecting the damping poles from
$1/\omega_{\rm P}$ and $1/\omega_{\rm H}$, the residual poles in $\delta\Delta_B(\omega)$
come from $\pm\omega_B$, leading to an oscillating behavior of 
$\delta\Delta_B(t)$ at frequency $\omega_B$ in the time domain.

{\em Rotating Higgs mode.---}We also find the optical
response of the rotating Higgs mode at weak scattering (for details, see Supplemental Material~\cite{SM})
\begin{equation}
[\omega^2_{R}(q)-\omega^2]\delta\Delta_R=\Big[\frac{(i{\bf q}\times{e{\bf
      E}_q})\cdot{\bf z}}{{m}}\Big](2\Delta_0)s_{\omega}, \label{RH}
\end{equation}
where  $s_{\omega}$ is the dimensionless response coefficient [see Supplemental Material, Eq.~(S33)~\cite{SM}];
the energy spectrum
\begin{equation}
\resizebox{.48\hsize}{!}{$\omega_{R}(q)=\sqrt{m^2_{R}+q^2v_F^2z/4}$},  
\end{equation}  
with
\begin{equation}
\resizebox{.62\hsize}{!}{$m_R^2=\frac{\Delta^2_0\sum_{\bf
      k}\sin^2(4\theta_{\bf k})[-\partial_{E_{\bf k}}f(E_{\bf k})]/E_{\bf k}^2}{\sum_{\bf
    k}\sin^2(2\theta_{\bf k})\xi^2_ka_{\bf
      k}/E^4_{\bf k}}$},
\end{equation}
and {\footnotesize $z=[{\sum_{\bf
    k}\sin^2(2\theta_{\bf k}){\xi_k}\partial^2_{\xi_k}(\xi_ka_{\bf k})/{E^2_{\bf
      k}}}]/[{\sum_{\bf
    k}\sin^2(2\theta_{\bf k}){\xi^2_{k}}a_{\bf
       k}/{E_{\bf k}^4}}]$}.

Then, it is found that  
the energy spectrum of the rotating Higgs mode $\omega_R$ exhibits
an excitation gap $m_{R}$ [{\small $m^2_R\propto|\partial_{E_{\bf k}}f(E_{\bf k})|$}]
in the long-wavelength limit, which vanishes at 
{\small $T=0~$K} but becomes finite at {\small $T\ne0$}.
This can be understood as follows. The rotating Higgs mode here is in
fact a Goldstone boson due to the spontaneous breaking of the
rotational symmetry by the formation of the $d$-wave order parameter~\cite{RG}, and hence, is
gapless at {\small $T=0~$K} according to the Goldstone theorem~\cite{Gm2}.
Whereas at finite temperature, due to the interaction with the thermally
excited Bogoliubov quasiparticle,
the rotating Higgs mode acquires 
a finite excitation gap, and thus, does not violate the
Mermin-Wagner theorem which rules out the long-range 
order (gapless excitation) at {\small $T\ne0$} in two-dimensional systems.
In addition, the energy dispersion of the rotating Higgs mode
exhibits the $s$-wave structure, which is because the collective
excitation in the long-wavelength limit is unaware of the pairing symmetry. 
The linear optical response, i.e., the right-hand side of Eq.~(\ref{RH}),
vanishes in the long-wavelength limit ($q=0$) as this mode is charge
neutral. Moreover, no second-order optical response is found 
in our calculation. Consequently, the rotating Higgs mode is in fact
optically invisible, in contrast to the breathing one.   

Actually, the finite response of the rotating Higgs mode, in principle, needs the
chiral-symmetry breaking, which can be achieved by the magnetic field.  
Specifically, in the pseudogap phase, considering the case with
magnetic field,  
we find the response of the rotating Higgs mode (for details, see Supplemental Material~\cite{SM})
\begin{equation}
  [\omega^2_{R}(q)-\omega^2]\delta\Delta_R=-\frac{eB_{z,q}}{m}\frac{\Delta^3_0}{i\omega_{\rm
  H}}c_1,
\end{equation}  
with $c_1$ being the response coefficient [see Supplemental Material, Eq.~(S34)~\cite{SM}].
Then, it is observed that the rotating Higgs mode responds to magnetic
field in the linear regime as expected, directly suggesting its possible detection 
by magnetic resonance measurement.

Finally, we show an interesting property of the rotating Higgs mode.
Inspired by recent experiment where a negative thermal Hall
signal contributed by unidentified charge-neutral excitation is discovered
in the pseudogap phase of cuprate superconductors~\cite{THCC}, we calculate the
thermal response of the rotating Higgs mode. Specifically, 
in the presence of both temperature gradient and magnetic field in the pseudogap phase, we find  (for details, see Supplemental Material~\cite{SM})
\begin{equation}
(\omega^2_{R}-\omega^2)\delta\Delta_R\!=\!\frac{\varepsilon_{q}({\bm \nabla}_RT\!\times\!i{\bf
q})\!\cdot\!{\bf
    z}}{2m}\frac{\Delta^2_0\partial_T\Delta_0}{i\omega_{\rm
    H}i\omega_{\rm P}}c_2-\frac{eB_{z}\Delta^3_0}{i\omega_{\rm
  H}m}c_1,\label{FRH}  
\end{equation}
with $c_2$ being the corresponding response coefficient [see Supplemental Material, Eq.~(S35)~\cite{SM}].
The first term on the right-hand side of above equation, by temperature
gradient of the order parameter $\partial_T\Delta_0$, 
provides a transverse drive effect on the
rotating-Higgs-mode excitation. 
Nevertheless, this term alone does not provide
any finite thermal current due to the chiral symmetry
[after chiral transformation $\delta\Delta_R\rightarrow-\delta\Delta_R$,
momentum ${\bf q}$ in Eq.~(\ref{FRH}) changes sign].
The magnetic field, i.e., second term on the
right-hand side, breaks this symmetry, and a finite thermal Hall
current can therefore be achieved. To calculate this current, we
define the bosonic field {\small $\phi_{\bf q}(t)=\delta\Delta_R(t,{\bf q})/D$}
with {\small $D$} being the scale parameter, and construct
its Lagrangian from the equation of motion [Eq.~(\ref{FRH})]: 
\begin{equation}
L=|\partial_t\phi_{\bf q}(t)|^2-\omega_R^2|\phi_{\bf q}(t)|^2+[M_{{\bf
    q}}\phi^*_{\bf q}(t)+h.c.].\label{LG}
\end{equation}
Here, $M_{\bf q}=F_{{\bf q},\omega_B}/D$ with $F_{{\bf q},\omega}$ labeling
the terms on the right-hand side of Eq.~(\ref{FRH}), in which we have
neglected the trivial damping pole by taking $\omega$ as $\omega_R$.
It is noted that the Lagrangian in Eq.~(\ref{LG}) is a typical
Klein-Gordon one in the field theory~\cite{FT}. Therefore, 
within the standard path integral method  (for details, see Supplemental Material~\cite{SM}), a finite thermal Hall 
current density, induced by the rotating Higgs mode, is directly derived: 
\begin{equation}\label{TC} 
{\bf j}_E=\frac{({\bm \nabla}_{\bf R}{T}\times{e}{\bf B})}{2mT}\big(\lambda_R\Delta_0\partial_T\Delta_0\big),
\end{equation}
where {\small
  $\lambda_R=\sum_{\bf
    q}\frac{zv^2_F}{4\omega_R}\frac{c_2c_2\varepsilon^2_{q}\Delta_0^4}{D^2(\Gamma^2_H+\omega_R^2)(\Gamma^2_p+\omega_R^2)}[\frac{2n_R+1}{2\omega_R}-\frac{\partial{n_R}}{\partial{\omega_R}}]$} with
$n_R=\frac{1}{e^{{\omega_R}/{T}}-1}$ being the Bose-distribution function.
We emphasize this thermal Hall current, induced by the
charge-neutral rotating Higgs mode, is totally different from the 
conventional thermal Hall current for electronic carriers which is 
generated by Lorentz force. The current here is generated due to 
temperature gradient of the order parameter and chiral-symmetry
breaking by magnetic field as mentioned above. Particularly, due to the negative sign in
$\partial_T\Delta_0$, the thermal Hall current here
possesses the opposite sign to the conventional one.

 \begin{figure}[htb]
   {\includegraphics[width=6.1cm]{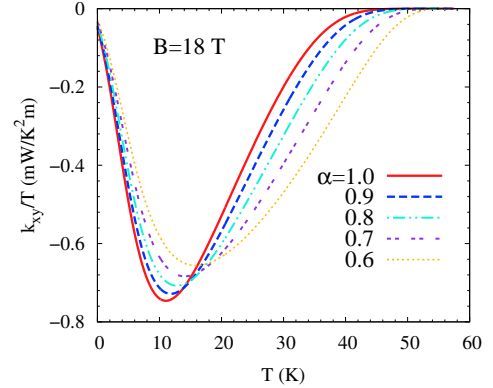}}
 \caption{(Color online) Temperature dependence of $k_{xy}/T$ from numerical
   calculation of Eq.~(\ref{TC}) with $\Delta_0=\gamma(T^*-T)^{\alpha}$.
   In the calculation, we choose a case in the overdoped regime which is far from the antiferromagnetic phase. $T^*=57~$K.
   Other used parameters can be found in the
   Supplemental Material (see Table SI \cite{SM}). }  
 \label{figyw1}
 \end{figure}

The temperature dependence of the thermal Hall conductivity
[$k_{xy}=j_E/(t\nabla_RT)$ with $t$ being the thickness of the single
  layer in cuprate superconductors]
requires the specific behavior of the ground state 
(i.e., pairing interaction) to obtain $\Delta_0(T)$, and this goes
beyond the scope of present analysis. 
Nevertheless, in Eq.~(\ref{TC}), with the increase of $T$ from 
  $0$ to $T^*$, $n_R$ increases, 
whereas the order parameter part {\small
  $\Delta_0\partial_T\Delta_0$}, in general, drops
slowly at the beginning and then rapidly near $T_c$. 
Thus, a peak behavior in the temperature dependence of $|k_{xy}|$ can
be expected. To justify this analysis, we approximately take
$\Delta_0\sim(T^*-T)^{\alpha}$ and perform a numerical calculation
of Eq.~(\ref{TC}). The numerical results for different $\alpha$
are plotted in
Fig.~\ref{figyw1}. As expected, with the increase of temperature from
$T=0~$K, $|k_{xy}/T|$ first increases and then decreases, leading to a
peak behavior observed around $T=12~$K. The experimental finding in the pseudogap phase
so far lies in the regime with $T{\ge}14~$K and only the decrease of
$|k_{xy}/T|$ is observed~\cite{THCC}. In this regime, the
experimentally observed linear dependence on magnetic field and temperature dependence
as well as negative sign of $k_{xy}/T$ in the pseudogap phase show good
agreement with our results. Therefore, we conjecture that the
experimentally observed unidentified charge-neutral excitation in
the pseudogap phase \cite{THCC} is the rotating Higgs mode. It is
noted that notwithstanding the fact that our computation in
Fig.~\ref{figyw1} extends to $T=0~$K, our result is valid only in
the pseudogap regime with $T_c<T<T^*$.

In conclusion, within the GIKE approach, we have analytically derived
the energy spectra of both breathing and rotating Higgs modes of the $d$-wave
order parameter for the first time. Then, investigations on their rich
dynamic properties have been carried out. Particularly, for 
the unique rotating Higgs mode in $d$-wave superconductors,
it is interesting to find that with longitudinal temperature
gradient in pseudogap phase, by magnetic field, this charge-neutral mode generates a
thermal Hall current, which is likely to capture
recent experimental finding in pseudogap phase
of cuprate superconductors~\cite{THCC}.

The authors acknowledge financial support from
the National Natural Science Foundation of 
China under Grants No.\ 11334014 and No.\ 61411136001.

\end{document}


\title{Theory of Higgs modes in $d$-wave superconductors \\ Supplement material}

\author{F. Yang}

\affiliation{Hefei National Laboratory for Physical Sciences at
Microscale, Department of Physics, and CAS Key Laboratory of Strongly-Coupled
Quantum Matter Physics, University of Science and Technology of China, Hefei,
Anhui, 230026, China}

\author{M. W. Wu}

\affiliation{Hefei National Laboratory for Physical Sciences at
Microscale, Department of Physics, and CAS Key Laboratory of Strongly-Coupled
Quantum Matter Physics, University of Science and Technology of China, Hefei,
Anhui, 230026, China}

\maketitle

\section{Gauge-invariant kinetic equation}

We first emphasize that by performing the gauge transformation {\small $\rho_{\bf
    k}(R)\rightarrow{e^{-i\tau_3\chi(R)}}\rho_{\bf
    k}(R)e^{i\tau_3\chi(R)}$},
the GIKE [Eq.~(4) in the main text] is gauge-invariant under the gauge
transformation first revealed by 
Nambu~\cite{gi0}:
\begin{eqnarray}
eA_{\mu}&\rightarrow&eA_{\mu}-\partial_{\mu}\chi(R), \label{gaugestructure1}\\
\delta\theta(R)&\rightarrow&\delta\theta(R)+2\chi(R),\label{gaugestructure2}
\end{eqnarray}
where the four vectors $A_{\mu}=(\phi,{\bf A})$ and
$\partial_{\mu}=(\partial_t,-{\bm \nabla}_{\bf R})$.  

Therefore, we can make the unitary transformation {\small $\rho_{\bf
    k}(R)\rightarrow{e^{i\tau_3\delta\theta(R)/2}}\rho_{\bf
    k}(R)e^{-i\tau_3\delta\theta(R)/2}$} mentioned in the main text,
and the GIKE becomes
\begin{eqnarray}
&&\partial_t\rho_{\bf
  k}\!+\!i\Big[\left(\xi_k\!+\!\mu_{\rm
      eff}\right)\tau_3\!+\!\Delta_{\bf k}\tau_1\!+\!\frac{\varepsilon_{\nabla_{\bf R}}}{4}\tau_3,\rho_{\bf 
      k}\Big]\!+\!\Big\{\frac{\bf k}{2m}\tau_3,{\bm \nabla}_{\bf R}\rho_{\bf k}\Big\}\!-\!\Big[\frac{{\bf
    p}_s\!\circ\!{\bm \nabla}_{\bf R}}{8m},\tau_3\rho_{\bf
    k}\Big]\!+\!\frac{1}{2}\left\{e{\bf E}\tau_3\!-\!({\bm
    \nabla}_{\bf R}\!+\!i{\bf p}_s\tau_3)\Delta_{\bf k}\tau_1,\partial_{\bf k}\rho_{\bf
k}\right\} \nonumber\\
&&\mbox{}\!-\!\frac{i}{8}\left[({\bm \nabla}_{\bf R}\!+\!i{\bf p}_s\tau_3)({\bm
    \nabla}_{\bf R}\!+\!i{\bf p}_s\tau_3)\Delta_{{\bf k}}\tau_1,\partial_{\bf k}\partial_{\bf k}\rho_{\bf
      k}\right]\!+\!\frac{{\bf k}\times{e}{\bf B}}{4m}[\tau_3,\tau_3\partial_{\bf k}\rho_{\bf
k}]\!+\!\frac{{\bf k}\!\times\!{e}{\bf B}}{2m}\{\tau_3,\tau_3\partial_{\bf k}\rho_{\bf
    k}\}
  =\!\partial_t\rho_{\bf k}\Big|_{\rm
scat},
\label{GE}
\end{eqnarray}
where {\small $\Delta_{\bf k}=\Delta_{{\bf k}0}+\delta\Delta_{\bf
    k}$}; the gauge-invariant {\small ${\bf p}_s={\bm \nabla}_{\bf
    R}\delta\theta-2e{\bf A}$} and {\small $\mu_{\rm
    eff}={\partial_t\delta\theta}/{2}+e\phi+\mu_{H}+{p^2_s}/({8m})$}.
It is noted that in Eq.~(\ref{GE}), the phase
fluctuation is effectively removed from the order parameter and manifests itself by ${\bf
  p}_s$ and $\mu_{\rm eff}$. Moreover, after the transformation, the equation of the order parameter [Eq.(3) in the main text] becomes
\begin{eqnarray}
{\sum_{\bf k'}}g_{\bf kk'}\rho_{{\bf k'}1}&=&-\Delta_{\bf k}\label{gap},\\
{\sum_{\bf k'}}g_{\bf kk'}\rho_{{\bf k'}2}&=&0\label{phase}.
\end{eqnarray}
Equation~(\ref{gap}) gives the gap equation and
hence the Higgs mode. Whereas Eq.~(\ref{phase}) determines the phase
fluctuation as revealed in our previous work~\cite{GOBE2}.
We emphasize that the calculation of the amplitude fluctuations is 
irrelevant of the phase fluctuation, since they represent mutually
orthogonal excitations and hence are decoupled.

It can be demonstrated that the charge conservation
is satisfied in the GIKE approach for
$d$-wave superconductivity in clean limit ($\partial_t\rho_{\bf
  k}|_{\rm scat}=0$). The gauge-invariant density and current
read~\cite{GOBE1,GOBE2} $n=\sum_{\bf k}(1+2\rho_{{\bf k}3})$ and ${\bf
  j}=\sum_{\bf k}\big(\frac{e{\bf k}}{m}\rho_{{\bf k}0}\big)$,
respectively. By taking the summation of the $\tau_3$ component of 
Eq.~(\ref{GE}) over ${\bf k}$, one has
\begin{equation}\label{T3}
\partial_{t}(\sum_{\bf k}\rho_{{\bf k}3})+{\bm \nabla_{\bf R}}\cdot\Big(\sum_{\bf k}\frac{\bf k}{m}\rho_{{\bf k}0}\Big)=\sum_{\bf k}2\Delta_{\bf k}\rho_{{\bf k}2}.  
\end{equation}
By using Eqs.~(\ref{gap}) and~(\ref{phase}), the right-hand side of
above equation {\small $\sum_{\bf k}2\Delta_{\bf k}\rho_{{\bf
      k}2}=-\sum_{\bf kk'}2(g_{\bf kk'}\rho_{{\bf k}'1})\rho_{{\bf
      k}2}=-\sum_{\bf kk'}\rho_{{\bf k}'1}(2g_{\bf k'k}\rho_{{\bf
      k}2})=0$} in which we have substituted $g_{\bf kk'}=g_{\bf
  k'k}$. Then, we immediately obtain the charge conservation {\small $\partial_t\delta{n}+{\bm \nabla_{\bf R}}\cdot{\bf j}=0$}. The charge conservation of the gauge-invariant kinetic theory is natural, since it has been proved long time ago by Nambu via the generalized Ward identity
that the gauge invariance in the superconducting states is equivalent to the charge conservation~\cite{gi0}.

\section{Scattering}

We next present the scattering terms $\partial_t\rho_{\bf k}|_{\rm
  scat}$ in Eq.~(\ref{GE}). Since the electron-phonon
scattering is weak at low temperature, we mainly consider the
electron-impurity scattering.
The specific impurity scattering terms, extending to $d$-wave
superconductivity from the $s$-wave case~\cite{GOBE3}, are written as
\begin{equation}
\partial_t\rho_{\bf k}\Big|_{\rm scat}=-[S_{\bf k}(>,<)-S_{\bf k}(<,>)+h.c.],\label{scat}
\end{equation}
with 
\begin{equation}
S_{\bf k}(>,<)=n_i\sum_{\bf k'}\int^{t}_{-\infty}dt'V_{\bf
    k-k'}\tau_3e^{i(t'-t){\hat H}_{\bf k'}}[1-\rho_{\bf k'}(t')]V_{\bf
    k'-k}\tau_3\rho_{\bf k}(t')e^{-i(t'-t){\hat H}_{\bf k}}.\label{SK}
\end{equation}
Here,  $n_i$ is the impurity density and 
$V_{\bf k-k'}$ stands for the electron-impurity interaction;
${\hat H}_{\bf k}=\xi_{{\bf
    k}+\tau_3{\bf p}_{s,i}}\tau_3+\Delta_{{\bf k}0}\tau_1$ denotes the
BCS-like Hamiltonian in the presence of the intrinsic phase fluctuation
$\delta\theta_{i}$. ${\bf p}_{s,i}={\bm \nabla}_{\bf
  R}\delta\theta_{i}$.

It is noted that this scattering term is
non-Markovian. In order to turn this non-Markovian scattering into
the Markovian one for further calculation, one needs to carefully handle the Markovian
approximation in a variety of situations of the external probe, as the 
previous work for $s$-wave superconductivity revealed~\cite{GOBE3}.
However, in high-$T_c$ superconductors, there exists large intrinsic phase
fluctuation $\delta\theta_i$ in the pseudogap phase. The specific
behaviors of this fluctuation is still unclear in the literature,
making it hard to rigorously calculate the scattering effect here. In the present analysis, we take the
relaxation-time approximation:
\begin{equation}
\partial_t\rho_{\bf k}\Big|_{\rm scat}=-\Gamma_p\delta\rho_{{\bf k}0}\tau_0-\Gamma_H\delta\rho_{{\bf k}-}\tau_--\Gamma_H\delta\rho_{{\bf k}+}\tau_+,
\end{equation}
in which the $\tau_3$ component of the scattering terms is ruled out
by the charge conservation in the 
impurity scattering. Here, $\Gamma_p$ ($\Gamma_H$) denotes the relaxation rate of 
the nonequilibrium $\delta\rho_{{\bf k}0}$ ($\delta\rho_{{\bf k}\pm}$).
We emphasize that in the main text, the scattering term/effect is
added only for completeness. It plays an insignificant role in our
results, since we focus on the excitation and
response to external probe.

\section{Solution of GIKE}

Considering a weak probe, we take the expansion {\small $\delta\rho_{\bf
    k}=\delta\rho^{(1)}_{\bf k}+\delta\rho^{(2)}_{\bf k}$}
with $\delta\rho^{(1)}_{\bf k}$ and $\delta\rho^{(2)}_{\bf k}$ being the first
and second order responses to the external probe.

\subsection{Linear response}

The first-order GIKE reads
\begin{eqnarray}
&&\partial_t\delta\rho^{(1)}_{\bf
  k}\!+\!i[\Xi_k\!\tau_3\!+\!\Delta_{{\bf k}0}\tau_1,\delta\rho^{(1)}_{\bf 
      k}]\!+\!i[\mu^{(1)}_{\rm
      eff}\tau_3\!+\!\delta\Delta^{(1)}_{\bf k}\tau_1,\rho^{(0)}_{\bf 
      k}]\!+\!\frac{1}{2}\{{\bf v_k}\tau_3,{\bm \nabla}_{\bf R}T\partial_T\rho^{(0)}_{\bf
    k}+{\bm \nabla}_{\bf R}\delta\rho^{(1)}_{\bf k}\}\!+\frac{1}{8m}\![{{\bm \nabla}_{\bf R}\!\cdot\!{\bf
      p}_s}\tau_3,\tau_3\rho^{(0)}_{\bf k}]
  \nonumber\\
  &&\mbox{}\!+\!\frac{1}{2}\{e{\bf
    E}\tau_3,\partial_{\bf k}\rho^{(0)}_{\bf k}\}\!-\!\frac{1}{2}\{{\bm \nabla}_{\bf R}T\partial_T\Delta_{{\bf
      k}0}\tau_1\!+\!{\bm \nabla}_{\bf R}\delta\Delta^{(1)}_{\bf k}\tau_1\!-\!{\bf
    p}_s\tau_2\Delta_{{\bf k}0},\partial_{\bf k}\rho^{(0)}_{\bf
    k}\}\!-\!\frac{i}{8}[{\bm \nabla}_{\bf R}{\bm \nabla}_{\bf R}\delta\Delta^{(1)}_{\bf
      k}\tau_1\!-\!{\bm \nabla}_{\bf R}{\bf p}_s\Delta_{{\bf
        k}0}\tau_2,\partial_{\bf k}\partial_{\bf k}\rho^{(0)}_{\bf
      k}]\nonumber\\
&&\mbox{}\!+\!\frac{1}{4}({\bf v_k}\!\times\!{\bf
    B})[\tau_3,\tau_3\partial_{\bf k}\rho^{(0)}_{\bf k}]\!+\!\frac{1}{2}({\bf v_k}\!\times\!{\bf
    B})\{\tau_3,\tau_3\partial_{\bf k}\rho^{(0)}_{\bf k}\}=\!-\Gamma_p\delta\rho^{(1)}_{{\bf k}0}\tau_0\!-\!\Gamma_H\delta\rho^{(1)}_{{\bf k}-}\tau_-\!-\!\Gamma_H\delta\rho^{(1)}_{{\bf k}+}\tau_+,\label{FO}
\end{eqnarray}
where $\Xi_k=\xi_k-\varepsilon_{\nabla_{\bf R}}/4$ and ${\bf v_k}={\bf
k}/m$; $\mu^{(1)}_{\rm eff}$
and $\delta\Delta^{(1)}_{\bf k}$ are corresponding first-order parts in $\mu_{\rm
  eff}$ and $\delta\Delta_{\bf k}$, respectively.
In CM frequency and momentum
space [{\small $R=(t,{\bf R})\rightarrow{q}=(\omega,{\bf q})$}], the
components of Eq.~(\ref{FO}) are given by
\begin{eqnarray}
&&(i\omega+\Gamma_p)\delta\rho^{(1)}_{{\bf k}0}=-{{\bf
      v_k}\!\cdot\!{\bm \nabla_{\bf R}T}}\partial_T\rho^{(0)}_{{\bf
      k}3}+{i{\bf v_k}\!\cdot\!{\bf q}}\delta\rho^{(1)}_{{\bf
      k}3}-i{\bf q}\!\cdot\!\partial_{\bf k}\rho^{(0)}_{{\bf
      k}1}\delta\Delta^{(1)}_{\bf k}+\partial_T\Delta_{{\bf
      k}0}{\bm \nabla}_{\bf R}T\!\cdot\!\partial_{\bf k}\rho^{(0)}_{{\bf
      k}1}-e{\bf E}_q\!\cdot\!\partial_{\bf k}\rho^{(0)}_{{\bf k}3},\label{11}\\
&&i\omega\delta\rho^{(1)}_{{\bf k}3}=2\Delta_{{\bf k}0}\delta\rho^{(1)}_{{\bf
      k}2}+{i{\bf v_k}\!\cdot\!{\bf q}}\delta\rho^{(1)}_{{\bf
      k}0}+{i}{\bf q}{\bf p}_{sq}\Delta_{{\bf
      k}0}/4:\partial_{\bf k}\partial_{\bf k}\rho^{(0)}_{{\bf
      k}1}-({{\bf v_k}\!\times\!{\bf B}_q})\!\cdot\!\partial_{\bf
    k}\rho^{(0)}_{{\bf k}3},\label{12}\\
&&(i\omega+\Gamma_H)\delta\rho^{(1)}_{{\bf k}1}=-2\Xi_k\delta\rho_{{\bf k}2}^{(1)}-{i}{\bf q}{\bf p}_{sq}\Delta_{{\bf
      k}0}/4:\partial_{\bf k}\partial_{\bf k}\rho^{(0)}_{{\bf
      k}3}-({{\bf v_k}\!\times\!{\bf B}_q})\!\cdot\!\partial_{\bf
    k}\rho^{(0)}_{{\bf k}1}/2+{i{\bf q}\!\cdot\!{\bf p}_{sq}}\rho^{(0)}_{{\bf k}1}/{(4m)},\label{13}\\
  &&(i\omega+\Gamma_H)\delta\rho^{(1)}_{{\bf k}2}=-2\Delta_{{\bf
      k}0}\delta\rho_{{\bf k}3}^{(1)}+2\Xi_k\delta\rho_{{\bf
      k}1}^{(1)}-2\rho^{(0)}_{{\bf k}3}\delta\Delta^{(1)}_{\bf k}+2\rho^{(0)}_{{\bf
      k}1}\mu^{(1)}_{\rm eff}-{\bf q}{\bf q}\delta\Delta^{(1)}_{{\bf
      k}}/4:\partial_{\bf k}\partial_{\bf k}\rho^{(0)}_{{\bf
      k}3}.\label{14}
\end{eqnarray}
Since the longitudinal vector potential does not
exist in either optical response or static magnetic response, 
one finds $i{\bf q}\cdot{\bf
  p}_{sq}=-q^2\delta\theta_q+i{\bf q}\cdot2e{\bf
  A}_q=-q^2\delta\theta_q$, which
in the long-wavelength limit can be neglected. 
Then, substituting Eqs.~(\ref{11})-(\ref{13}) into Eq.~(\ref{14}),
one has
\begin{eqnarray}
&&[4\Xi^2_k+4(\Delta_{{\bf
      k}0})^2+(i\omega_{\rm H})^2]\delta\rho^{(1)}_{{\bf
    k}2}+2\Delta_{{\bf k}0}\delta\rho^{(1)}_{{\bf
    k}3}\Big[\Gamma_H+\frac{({{\bf v_k}\!\cdot\!{\bf
        q}})^2}{\omega_{\rm P}}\Big]=-\frac{2\Delta_{{\bf k}0}}{\omega_{\rm P}}({{\bf v_k}\!\cdot\!{\bf
        q}})[\partial_T\Delta_{{\bf
      k}0}{\bm \nabla}_{\bf R}T\!\cdot\partial_{\bf k}\rho^{(0)}_{{\bf
      k}1}-{{\bf
      v_k}\!\cdot\!{\bm \nabla_{\bf R}T}}\partial_T\rho^{(0)}_{{\bf
        k}3}\nonumber\\
&&\mbox{}-i({\bf q}\cdot\partial_{\bf k})\rho^{(0)}_{{\bf
      k}1}\delta\Delta^{(1)}_{\bf k}-e{\bf E}_q\!\cdot\!\partial_{\bf
      k}\rho^{(0)}_{{\bf k}3}]-\frac{\Delta_{{\bf k}0}}{2}i{\bf q}{\bf p}_{sq}:(\Delta_{{\bf k}0}\partial_{\bf
    k}\partial_{\bf k}\rho^{(0)}_{{\bf k}1}+\Xi_k\partial_{\bf
    k}\partial_{\bf k}\rho^{(0)}_{{\bf k}3})+({\bf v_k}\!\times\!{\bf
    B})\cdot(2\Delta_{{\bf k}0}\partial_{\bf k}\rho^{(0)}_{{\bf
      k}3}-\Xi_{\bf k}\partial_{\bf k}\rho^{(0)}_{{\bf k}1})\nonumber\\
&&\mbox{}-2\rho^{(0)}_{{\bf k}3}i\omega_{\rm H}\delta\Delta^{(1)}_{\bf k}+2\rho^{(0)}_{{\bf
      k}1}i\omega_{\rm H}\mu^{(1)}_{\rm eff}-\frac{1}{4}{\bf q}{\bf q}i\omega_{\rm H}\delta\Delta^{(1)}_{{\bf
      k}}:\partial_{\bf k}\partial_{\bf k}\rho^{(0)}_{{\bf
      k}3}.
\end{eqnarray}
Considering the weak scattering and long-wavelength case
($\Delta_0>\Gamma_H,v_kq$), we neglect the scattering and CM momentum
part on the left-hand side of the above equation.
Then, one immediately obtains the solution of $\delta\rho^{(1)}_{{\bf k}2}$,
from which $\delta\rho^{(1)}_{{\bf k}1}$ by Eq.~(\ref{13}) is derived:
\begin{equation}
i\omega_H\delta\rho^{(1)}_{{\bf
      k}1}=-i\omega_H\delta\Delta^{(1)}_{\bf k}H^{(0)}_{\bf
  k}+E^{(1)}_{\bf k}-P^{(1)}_{\bf k}.
\end{equation}
Here, $P^{(1)}_{\bf k}=4\Xi_kz_{\omega,{\bf k}}i\omega_H\rho^{(0)}_{{\bf
    k}1}\mu^{(1)}_{\rm eff}$; $H^{(0)}_{\bf k}$ and $E^{(1)}_{\bf k}$ read
\begin{eqnarray}
H^{(0)}_{\bf k}&=&-4\Xi_kz_{\omega,{\bf k}}\rho^{(0)}_{{\bf k}3}\!-\!{\Xi_kz_{\omega,{\bf k}}}{\bf q}{\bf q}:\partial_{\bf k}\partial_{\bf k}\rho^{(0)}_{{\bf
      k}3}/2\!-\!{4\Delta_{{\bf k}0}\Xi_kz_{\omega,{\bf k}}}({{\bf v_k}\!\cdot\!{\bf
        q}})({\bf q}\cdot\partial_{\bf k})\rho^{(0)}_{{\bf
    k}1}/({i\omega_{\rm P}i\omega_H}),\\
E^{(1)}_{\bf k}&=&\Xi_kz_{\omega,{\bf k}}\Big\{{4\Delta_{{\bf k}0}}({{\bf v_k}\!\cdot\!{\bf
        q}})/{\omega_{\rm P}}[{\bm \nabla}_{\bf R}T\!\cdot(\partial_T\Delta_{{\bf
      k}0}\partial_{\bf k}\rho^{(0)}_{{\bf
      k}1}\!-\!{\bf
      v_k}\partial_T\rho^{(0)}_{{\bf
        k}3})-\!e{\bf E}_q\!\cdot\!\partial_{\bf
      k}\rho^{(0)}_{{\bf k}3}]\!+\!i{\bf q}{\bf
  p}_{sq}\Delta_{{\bf k}0}:(\Delta_{{\bf k}0}\partial_{\bf
    k}\partial_{\bf k}\rho^{(0)}_{{\bf k}1}\!\nonumber\\
&&\mbox{}\!+\!\Xi_k\partial_{\bf
    k}\partial_{\bf k}\rho^{(0)}_{{\bf k}3})\!-\!2({\bf v_k}\!\times\!{\bf
    B})\cdot(2\Delta_{{\bf k}0}\partial_{\bf k}\rho^{(0)}_{{\bf
      k}3}\!-\!\Xi_{\bf k}\partial_{\bf k}\rho^{(0)}_{{\bf k}1})\Big\}-{i}{\bf q}{\bf p}_{sq}\Delta_{{\bf
      k}0}/4:\partial_{\bf k}\partial_{\bf k}\rho^{(0)}_{{\bf
      k}3}-({{\bf v_k}\!\times\!{\bf B}_q})\!\cdot\!\partial_{\bf
    k}\rho^{(0)}_{{\bf k}1}/{2}.~~~
\end{eqnarray}
Finally, from Eqs.~(8) and ~(9) in the main  text, the breathing
and rotating Higgs modes are derived:
\begin{eqnarray}
&&i\omega_H\delta\Delta^{(1)}_B/D_2=-i\omega_H\sum_{\bf k}\cos(2\theta_{\bf k})\delta\rho^{(1)}_{{\bf
      k}1}=i\omega_H\delta\Delta^{(1)}_{B}\sum_{\bf k}\cos^2(2\theta_{\bf k})H^{(0)}_{\bf
  k}-\sum_{\bf k}\cos(2\theta_{\bf k})E^{(1)}_{\bf k},\label{FB}\\
&&i\omega_H\delta\Delta^{(1)}_R/D_2=-i\omega_H\sum_{\bf k}\sin(2\theta_{\bf k})\delta\rho^{(1)}_{{\bf
      k}1}=i\omega_H\delta\Delta^{(1)}_{R}\sum_{\bf k}\sin^2(2\theta_{\bf k})H^{(0)}_{\bf
  k}-\sum_{\bf k}\sin(2\theta_{\bf k})E^{(1)}_{\bf k},\label{FR} 
\end{eqnarray}
where we have taken care of
the particle-hole symmetry to remove terms with the
odd order of $\xi_k$ in the summation of ${\bf k}$.
From Eq.~(\ref{gap}) in equilibrium state, one has
\begin{equation}
\frac{1}{D_2}=\sum_{\bf k}\cos^2(2\theta_{\bf k})\frac{1-2f(E_{\bf k})}{2E_{\bf k}},\label{DB}
\end{equation}  
and its equivalent variation [see Eq.~(\ref{EV}) in the following section] 
\begin{equation}
\frac{1}{D_2}=\sum_{\bf k}\sin^2(2\theta_{\bf
  k})\Big[\frac{\xi^2_k}{E^2_{\bf k}}\frac{1-2f(E_{\bf k})}{2E_{\bf
    k}}-\frac{\Delta^2_{{\bf k}0}}{E_{\bf k}^2}\partial_{E_{\bf k}}f(E_{\bf k})\Big].\label{DR}
\end{equation}
Substituting Eqs.~(\ref{DB}) and (\ref{DR}) into Eqs.~(\ref{FB}) and
(\ref{FR}) respectively, 
we obtain the linear responses of the breathing
and rotating Higgs modes to the external probe in the main text.
Particularly, for optical response, $E\ne0$, $A\ne0$, $B\approx0$ and
$\nabla_RT=0$; for magnetic response, $E=0$, $A\ne0$, $B\ne0$ and
$\nabla_RT=0$; for thermal-Hall investigation, $E=0$, $A\ne0$,
$B\ne0$ and $\nabla_RT\ne0$.

\subsection{Second-order response}

In second-order regime, we only focus on the optical response in
long-wavelength limit $(q=0)$ where $\delta\Delta^{(1)}_{\bf k}$ vanishes.
Then, the second-order GIKE reads
\begin{equation}
\partial_t\delta\rho^{(2)}_{\bf
  k}\!+\!i[\xi_k\!\tau_3\!+\!\Delta_{{\bf k}0}\tau_1,\delta\rho^{(2)}_{\bf 
      k}]\!+\!i[\mu^{(2)}_{\rm
      eff}\tau_3\!+\!\delta\Delta^{(2)}_{\bf k}\tau_1,\rho^{(0)}_{\bf 
      k}]\!+\!\frac{1}{2}\{e{\bf
    E}\tau_3\!+\!{\bf
    p}_s\Delta_{{\bf k}0}\tau_2,\partial_{\bf k}\rho^{(1)}_{\bf
    k}\}\!+\!\frac{i}{8}[{\bf p}_s{\bf p}_s\Delta_{{\bf
      k}0}\tau_1,\partial_{\bf k}\partial_{\bf k}\rho^{(0)}_{\bf
      k}]=\partial_t\rho_{\bf k}|^{(2)}_{\rm scat},\label{SO}
\end{equation}
where $\partial_t\rho_{\bf k}|^{(2)}_{\rm scat}=-\Gamma_p\delta\rho^{(2)}_{{\bf
    k}0}\tau_0\!-\!\Gamma_H\delta\rho^{(2)}_{{\bf
    k}-}\tau_-\!-\!\Gamma_H\delta\rho_{{\bf k}+}^{(2)}\tau_+$; $\mu^{(2)}_{\rm eff}$
and $\delta\Delta^{(2)}_{\bf k}$ are corresponding second-order parts in $\mu_{\rm
  eff}$ and $\delta\Delta_{\bf k}$, respectively. 
In CM frequency and momentum space [{\small $R=(t,{\bf
      R})\rightarrow{q}=(\omega,{\bf q})$}], with the linear
solution {\small $\delta\rho_{\bf k}^{(1)}=-[e{\bf E}_q\!\cdot\!\partial_{\bf
  k}\rho^{(0)}_{{\bf k}3}/(i\omega_{\rm P})]\tau_0$} in the long-wavelength limit
for optical response [see Eqs.~(\ref{11})-(\ref{14})], the components
of Eq.~(\ref{SO}) are given by 
\begin{eqnarray}
&&i\omega_{\rm P}\delta\rho^{(2)}_{{\bf k}0}=0,\label{21}\\
&&i\omega\delta\rho^{(2)}_{{\bf k}3}=2\Delta_{{\bf k}0}\delta\rho^{(2)}_{{\bf
      k}2}-\overline{e{\bf E}_q\!\cdot\!\partial_{\bf k}\delta\rho_{{\bf k}0}^{(1)}},\label{22}\\
&&i\omega_{\rm H}\delta\rho^{(2)}_{{\bf k}1}=-2\xi_k\delta\rho^{(2)}_{{\bf k}2},\label{23}\\
&&i\omega_{\rm H}\delta\rho^{(2)}_{{\bf k}2}=2(\xi_k\delta\rho_{{\bf
      k}1}^{(2)}-\Delta_{{\bf
      k}0}\delta\rho_{{\bf k}3}^{(2)}-\rho^{(0)}_{{\bf k}3}\delta\Delta^{(2)}_{\bf k}+\rho^{(0)}_{{\bf
      k}1}\mu^{(2)}_{\rm eff})-\overline{{\bf p}_{sq}{\bf p}_{sq}}\Delta_{{\bf
      k}0}/4:\partial_{\bf k}\partial_{\bf k}\rho^{(0)}_{{\bf
      k}3}-\Delta_{{\bf k}0}\overline{{\bf p}_{sq}\!\cdot\!\partial_{\bf k}\delta\rho_{{\bf k}0}^{(1)}}.\label{24}
\end{eqnarray}
In analogy to above derivation of linear response, one can straightforwardly
obtain the solution of $\delta\rho^{(2)}_{{\bf k}2}$:
\begin{equation}
\delta\rho^{(2)}_{{\bf k}1}=-2\xi_kz_{\omega,{\bf k}}(2\rho^{(0)}_{{\bf
      k}1}\mu^{(2)}_{\rm eff}-2\rho^{(0)}_{{\bf k}3}\delta\Delta^{(2)}_{\bf
  k})-2\xi_kY^{(2)}_{\bf k}, 
\end{equation}
with $Y^{(2)}_{\bf k}=z_{\omega,{\bf k}}\Delta_{{\bf k}0}\big(\frac{2}{i\omega_{\rm H}}\overline{e{\bf
    E}_q\!\cdot\!\partial_{\bf k}\delta\rho_{{\bf
      k}0}^{(1)}}-\overline{{\bf A}_{q}{\bf A}_{q}:\partial_{\bf k}\partial_{\bf k}\rho^{(0)}_{{\bf
      k}3}}+2\overline{{\bf A}_{q}\!\cdot\!\partial_{\bf k}\delta\rho_{{\bf k}0}^{(1)}}\big)$.

From Eq.~(\ref{23}) as well as Eqs.~(8) and ~(9) in the main text, the breathing
and rotating Higgs modes are derived:
\begin{eqnarray}
&&\delta\Delta^{(2)}_B/D_2=-\sum_{\bf k}\cos(2\theta_{\bf k})\delta\rho^{(2)}_{{\bf
      k}1}=-\delta\Delta^{(2)}_{B}\sum_{\bf k}\cos^2(2\theta_{\bf
    k})4\xi_kz_{\omega,{\bf k}}\rho^{(0)}_{{\bf k}3}+\sum_{\bf k}\cos(2\theta_{\bf k})2\xi_kY^{(2)}_{\bf k},\label{SB}\\
&&\delta\Delta^{(2)}_R/D_2=-\sum_{\bf k}\sin(2\theta_{\bf k})\delta\rho^{(2)}_{{\bf
      k}1}=-\delta\Delta^{(2)}_{R}\sum_{\bf k}\sin^2(2\theta_{\bf k})4\xi_kz_{\omega,{\bf k}}\rho^{(0)}_{{\bf k}3}+\sum_{\bf k}\sin(2\theta_{\bf k})2\xi_kY^{(2)}_{\bf k},\label{SR} 
\end{eqnarray}
where we have taken care of
the particle-hole symmetry to remove terms with the
odd order of $\xi_k$ in the summation of ${\bf k}$. 
Substituting Eqs.~(\ref{DB}) and (\ref{DR}) into Eqs.~(\ref{SB}) and
(\ref{SR}) respectively, 
we obtain the second-order optical responses of the breathing and
rotating Higgs modes. Particularly, the last term in Eq.~(\ref{SR})
is zero after the summation of $\theta_{\bf k}$, indicating the vanishing
second-order optical response of the rotating Higgs mode.

The corresponding response coefficients in the main text are given by
\begin{eqnarray}
u_{\omega}&=&\frac{\sum_{\bf
    k}\xi_k^2\cos^2(2\theta_{\bf
    k})z_{\omega,{\bf k}}\partial_{\xi_k}(\xi_ka_{\bf k})}{\omega_{\rm
    P}\omega_{\rm H}\sum_{\bf k}\cos^2(2\theta_{\bf
    k})z_{\omega,{\bf k}}a_{\bf k}},\\
d_{\omega}&=&\frac{\sum_{\bf
    k}\xi_k\cos^2(2\theta_{\bf
    k})z_{\omega,{\bf k}}\partial^2_{\xi_k}(\xi_ka_{\bf k})}{\sum_{\bf k}\cos^2(2\theta_{\bf
    k})a_{\bf k}z_{\omega,{\bf k}}},\\
s_{\omega}&=&\frac{\Delta_0^2\sum_{\bf
    k}\xi^2_k/E_k\sin^2(4\theta_{\bf k})z_{\omega,{\bf
      k}}\partial_{E_{\bf k}}a_{\bf k}}{i\omega_{\rm P}i\omega_{\rm H}\sum_{\bf
    k}\xi^2_k/E^2_{\bf k}\sin^2(2\theta_{\bf k})z_{\omega,{\bf k}}a_{\bf
    k}},\\
c_1&=&\frac{\sum_{\bf
    k}\sin^2(4\theta_{\bf k})(\xi_k/E^2_{\bf k})^2\big[-2\partial_{E_{\bf
        k}}f(E_{\bf k})\big]}{\sum_{\bf
    k}\sin^2(2\theta_{\bf k})({\xi_k}/{E^2_{\bf k}})^2a_{\bf
    k}},\\
c_2&=&\frac{\sum_{\bf k}\sin^2(4\theta_{\bf
    k})(1/E_{\bf k})^2\partial_{\Delta_{{\bf
        k}0}}(\Delta_{{\bf k}0}a_{\bf k})}{\sum_{\bf
      k}2\sin^2(2\theta_{\bf k})(\xi_k/E^2_{\bf k})^2a_{\bf k}}.
\end{eqnarray}

\section{Rotational Symmetry}

Equation~(\ref{gap}) in equilibrium state reads
\begin{equation}
{\Delta_{{\bf k}0}}=\sum_{\bf k'}g_{\bf kk'}\frac{\Delta_{{\bf k'}0}}{E_{\bf k'}}\frac{1-2f(E_{\bf k'})}{2},\label{RS}
\end{equation}
which exhibits the rotational symmetry of the polar axis of ${\bf k}$.
Then, in the above equation, by first replacing $\theta_{\bf k}$ and
$\theta_{\bf k'}$ with $\theta_{\bf k}-\alpha$ and $\theta_{\bf
  k'}-\alpha$ and then taking the derivative of $\alpha$ on both
sides of the equation, one obtains an equivalent variation of
Eq.~(\ref{RS}) after setting $\alpha=0$:
\begin{equation}
\Delta_0\sin(2\theta_{\bf k})=\sum_{\bf k'}g_{\bf kk'}\Delta_0\sin(2\theta_{\bf k'})\Big[\frac{\xi^2_{k'}}{E^2_{\bf k'}}\frac{1-2f(E_{\bf k'})}{2E_{\bf
    k'}}-\frac{\Delta^2_{{\bf k'}0}}{E_{\bf k'}^2}\partial_{E_{\bf
    k'}}f(E_{\bf k'})\Big]. \label{EV}  
\end{equation}

\section{Thermal Hall Current}

From the Lagrangian of the bosonic field $\phi_{\bf q}(t)$ [Eq.~(18) in
  the main text], one has the action 
\begin{equation}\label{action}
S=\int{dt}\int{d{\bf
    q}}L(\phi_{\bf q},\phi^*_{\bf q},\partial_t\phi_{\bf q},\partial_t\phi^*_{\bf q}).
\end{equation}
Considering the variation $\delta_{\phi^*_{\bf q}}{S}=0$, one obtains
\begin{equation}
0=\int{dt}\int{d{\bf q}}\Big[\frac{\partial{L}}{\partial\phi^*_{\bf q}}\delta\phi^*_{\bf q}+\frac{\partial{L}}{\partial(\partial_t\phi^*_{\bf q})}\delta(\partial_t\phi^*_{\bf q})\Big]=\int{dt}\int{d{\bf q}}\Big\{\frac{\partial{L}}{\partial\phi^*_{\bf q}}\delta\phi^*_{\bf q}-\partial_t\Big[\frac{\partial{L}}{\partial(\partial_t\phi^*_{\bf q})}\Big]\delta\phi^*_{\bf q}+\partial_t\Big[\frac{\partial{L}}{\partial(\partial_t\phi^*_{\bf q})}\delta\phi^*_{\bf q}\Big]\Big\}.  
\end{equation}
The last term vanishes since $\delta\phi_{\bf q}$ is zero at temporal
beginning and end. Then, we arrive the Euler-Lagrange equation of
motion
$\partial_t\Big[\frac{\partial{L}}{\partial(\partial_t\phi^*_{\bf q})}\Big]-\frac{\partial{L}}{\partial\phi^*_{\bf q}}=0$.
Substituting the specific Lagrangian [Eq.~(18) in
  the main text] into this equation, one recovers Eq.~(17) in the main text. 

To calculate the thermal property of the bosonic field, we map the action
in Eq.~(\ref{action}) into the imaginary-time one~\cite{G1} and then obtain
\begin{equation}
S=\int^{\beta}_0d\tau\int{d{\bf
    q}}[\phi^*_{\bf q}(\tau)D^{-1}({\tau},q)\phi_{\bf q}(\tau)+M^*_{\bf
    q}\phi_{\bf q}(\tau)+M_{\bf
    q}\phi^*_{\bf q}(\tau)],  
\end{equation}
where $\beta=1/T$ and the Green function
$D({\tau},q)=[\partial_{\tau}^2-\omega_R^2(q)]^{-1}$. Then, following the
standard generating functional method within the path integral approach~\cite{FT},
the thermal current density is written as 
\begin{eqnarray}
{\bf j}_E&=&T/Z_0\int^{\beta}_0d{\tau'}\sum_{\bf q'}\omega_R(q'){\bf
  v}_{\bf q'}\langle|\phi_{\bf q'}^*(\tau')\phi_{\bf q'}(\tau'){e^{-S}}|\rangle \nonumber\\
&=&T/Z_0\int^{\beta}_0d{\tau'}\sum_{\bf q'}\omega_R(q'){\bf
  v}_{\bf q'}\langle|\delta_{J^*_{\bf q'}}\delta_{J_{\bf q'}}e^{-\int^{\beta}_0d\tau\int{d{\bf
    q}}[\phi^*_{\bf q}(\tau)D^{-1}({\tau},q)\phi_{\bf q}(\tau)-{\bar{J}}_{\bf
    q}\phi_{\bf q}(\tau)-{\bar{J}}^*_{\bf
      q}\phi^*_{\bf q}(\tau)]}|\rangle\big|_{J=J^*=0}\nonumber\\
&=&T/Z_0\int^{\beta}_0d{\tau'}\sum_{\bf q'}\omega_R(q'){\bf
  v}_{\bf q'}\delta_{J^*_{\bf q'}}\delta_{J_{\bf
    q'}}\Big\{\int{D}\phi_{\bf q}D\phi^*_{\bf q}e^{-\int^{\beta}_0d\tau\int{d{\bf
    q}}[\phi^*_{\bf q}(\tau)D^{-1}({\tau},q)\phi_{\bf q}(\tau)-{\bar{J}}_{\bf
    q}D(\tau,q){\bar{J}}^*_{\bf
      q}]}\Big\}\Big|_{J=J^*=0}\nonumber\\
&=&T\int^{\beta}_0d{\tau'}\sum_{\bf q'}\omega_R(q'){\bf
  v}_{\bf q'}\delta_{J^*_{\bf q'}}\delta_{J_{\bf
    q'}}e^{{\rm Tr}[{\bar{J}}_{\bf
    q}D(\tau,q){\bar{J}}^*_{\bf
        q}]}\big|_{J=J^*=0}\big[e^{-{\rm
    Tr}\{\ln[D^{-1}(\tau,q)]\}}/Z_0\big]\nonumber\\
&=&T\int^{\beta}_0d{\tau'}\sum_{\bf q'}\omega_R(q'){\bf
  v}_{\bf q'}\delta_{J^*_{\bf q'}}\delta_{J_{\bf
    q'}}e^{{\rm Tr}[{\bar{J}}_{\bf
    q}D(\tau,q){\bar{J}}^*_{\bf
        q}]}\big|_{J=J^*=0}.\label{TCD}
\end{eqnarray}
where ${\bar{J}}_{\bf q}=J_{\bf q}-M^*_{\bf q}$ with $J_{\bf q}$ being
the generating functional; $\delta_{J_{\bf q}}$ denotes the functional
derivative; $Z_0$ stands for the partition function and ${\bf v}_{\bf
  q}=\partial_{\bf q}\omega_R(q)$ represents the group velocity.
In Matsubara frequency space, considering the weak external probe, the above equation becomes
\begin{eqnarray}
{\bf j}_E&=&\sum_{n,{\bf q}}\frac{\omega_R{\bf
  v}_{\bf q}M^*_{\bf q}M_{\bf
    q}}{[(i\omega_n)^2-\omega^2_R]^2}\exp\Big[\frac{1}{\beta}\sum_{n',{\bf
  q'}}\frac{M^*_{\bf q'}M_{\bf
      q'}}{(i\omega_{n'})^2-\omega^2_R}\Big]\approx\sum_{n,{\bf q}}\frac{\omega_R{\bf
  v}_{\bf q}M^*_{\bf q}M_{\bf
    q}}{[(i\omega_n)^2-\omega^2_R]^2}\nonumber\\
&=&\frac{({\bm \nabla}_{\bf R}{T}\times{e}{\bf B})}{mT}\big(\frac{\Delta_0}{D}\big)^2\frac{\partial_T\Delta_0}{2}\sum_{\bf
  q}\frac{c_1c_2zv^2_F\varepsilon^2_{q}}{4\Delta_0}\frac{\Delta_0^2}{\Gamma^2_H+\omega_R^2}\frac{\Delta_0^2}{\Gamma^2_p+\omega_R^2}T\sum_n\frac{2\omega_R}{[(i\omega_n)^2-\omega^2_R]^2}.
\end{eqnarray}
Then, after the summation of the Matsubara frequency~\cite{G1}, Eq.~(19) in the
main text is obtained.

\begin{table}[htb]
  \caption{Parameters used in the numerical calculation of
    Eq.~(19) in the main text. Noted that $m_0$ stands for the free
    electron mass. $\gamma$ is obtained by fitting
    $E_0=\Delta_0(T=0)$. $D$ reflects the amplitude of the fluctuation
    $\delta\Delta_R$. We choose the case in the overdoped regime
    which is far from the antiferromagnetic phase. } 
\label{Table}
  \begin{tabular}{l l l l l l l l}
    \hline
    \hline   
    $m/m_0$&\;\;\;\;$5^a$&\;\;\;$E_F~$(meV)&\;\;\;\;$50^b$&\;\;\;$E_0~$(meV)&\;\;\;\;$2^c$&\;\;\;$T^*~$(K)&\;\;\;\;$57^d$\\
    $\Gamma_p/E_0$&\;\;\;\;$0.2$&\;\;\;$\Gamma_H/E_0$&\;\;\;\;$0.1$&\;\;\;$D/E_0$&\;\;\;\;$0.02$&\;\;\;$t~$(nm)&\;\;\;\;$1.3^e$\\
    \hline
    \hline
  \end{tabular}\\
  \quad$^a$ Ref.~\onlinecite{parameter1}. \quad$^b$
  Ref.~\onlinecite{parameter2}. \quad$^c$
  Ref.~\onlinecite{parameter3}. \quad$^d$ Ref.~\onlinecite{THCC}. \quad$^e$ Ref.~\onlinecite{TN}.
  
\end{table}